\documentclass[a4paper]{jpconf}
\usepackage{graphicx}
\newcommand{\ba}{\begin{eqnarray}}
\newcommand{\ea}{\end{eqnarray}}

\begin{document}

\title{Heavy $\Omega_c$ and $\Omega_b$ baryons in the quark model}

\author{E. Ortiz-Pacheco and R. Bijker}
\address{Instituto de Ciencias Nucleares, 
Universidad Nacional Aut\'onoma de M\'exico, 
A.P. 70-543, 04510 Ciudad de M\'exico, M\'exico}
\ead{bijker@nucleares.unam.mx}
\author{A. Giachino and E. Santopinto}
\address{INFN, Sezione di Genova, Via Dodecaneso 33, 16146 Genova, Italy}

\begin{abstract}
In this contribution, we present a study of ground- and excited-state $\Omega_c$ 
and $\Omega_b$ baryons consisting of two strange quarks and a heavy charm or bottom quark. An analysis in the quark model shows that the recently observed excited $\Omega_c$ and $\Omega_b$ states can be interpreted in terms of $\lambda$-mode excitations. 
\end{abstract}

\section{Introduction}

In recent years there has been a renewed interest in hadron physics especially concerning 
hadrons containing heavy (charm or bottom) quarks. The experimental discovery of many new 
heavy baryons \cite{PDG} as well as candidates for multiquark configurations like tetraquark 
and pentaquark states \cite{review1,review2} has sparked a large number of studies into the 
structure of hadrons. 

In particular, we mention the discovery of five new $\Omega_c$ states by the LHCb Collaboration 
in the $\Xi_c^+ K^-$  decay channel \cite{Omegac} and the subsequent confirmation of four of these 
states by the Belle Collaboration \cite{Belle}: $\Omega_c(3000)$, $\Omega_c(3050)$, $\Omega_c(3065)$ 
and $\Omega_c(3090)$. In addition, the $\Omega_c(3188)$, even if not yet confirmed as a genuine 
resonance for lack of sufficient statistical significance, was seen by both LHCb and Belle, whereas 
the $\Omega_c(3119)$ was observed only by LHCb, but not by Belle. 

Since neither LHCb nor Belle were able to determine angular momenta and parities, 
the assignment of quantum numbers is model dependent. Several different assignments exist in the 
literature, see {\it e.g.} \cite{Karliner:2017kfm,Zhao:2017fov,Wang:2017hej,Padmanath:2017lng,
Agaev:2017lip,epjc}. In particular, for the $\Omega_c(3119)$ there exists a variety of spin and 
parity assignments: $J^{P}=\frac{1}{2}^{+}$ or $\frac{3}{2}^{+}$ \cite{Wang:2017hej}, $\frac{5}{2}^{-}$ 
\cite{Padmanath:2017lng}, $\frac{3}{2}^{+}$ or $\frac{5}{2}^{-}$ \cite{Karliner:2017kfm}. 

The aim of this contribution is to present a quark model study of ground- and excited-state 
$\Omega_c$ baryons, and to show that they can be interpreted in terms of $\lambda$-mode excitations. 
The analysis is based both on masses and decay widths. A similar study is carried out for $\Omega_b$ 
baryons which subsequently was confirmed by new experimental data from the LHCb Collaboration 
\cite{Omegab}.

\section{$\Omega_c$ and $\Omega_b$ baryons}

In the quark model, $\Omega_c$ and $\Omega_b$ baryons correspond to $ssQ$ configurations 
consisting of two strange quarks and one heavy quark, $Q=c$ and $b$, respectively. 
In this contribution we consider a harmonic oscillator quark model with a spin, spin-orbit, 
isospin and flavor dependent terms according to Ref.~\cite{epjc}
\ba
H \;=\; H_{\rm ho} + A \, \vec{S} \cdot \vec{S} + B \, \vec{L} \cdot \vec{S} 
+ E \, \vec{I} \cdot \vec{I} + F \, C_{2SU_{\rm f}(3)} ~.
\label{hosc}
\ea
The harmonic oscillator quark model for $ssQ$ baryons with two equal masses 
$m_1=m_2=m_s$ different from the third $m_3=m_Q$ is given by \cite{IK}
\ba
H_{\rm ho} &=& \sum_i \left( m_i + \frac{p_i^2}{2m_i} \right) 
+ \frac{1}{2} C \sum_{i<j} (\vec{r}_i - \vec{r}_j)^2 
\nonumber\\ 
&=& 2m_s+m_Q + \frac{P^2}{2(2m_s+m_Q)} 
+ \frac{p_{\rho}^2}{2m_{\rho}} + \frac{p_{\lambda}^2}{2m_{\lambda}} 
+ \frac{1}{2} m_{\rho} \omega_{\rho}^2 \rho^2 
+ \frac{1}{2} m_{\lambda} \omega_{\lambda}^2 \lambda^2 ~,
\ea
where we have made a change of variables to relative Jacobi coordinates and 
the center-of-mass coordinate
\ba
\vec{\rho} &=& (\vec{r}_1 - \vec{r}_2)/\sqrt{2} ~,
\nonumber\\
\vec{\lambda} &=& (\vec{r}_1 + \vec{r}_2 - 2\vec{r}_3)/\sqrt{6} ~,
\nonumber\\
\vec{R} &=& \frac{m_s(\vec{r}_1 + \vec{r}_2) + m_Q\vec{r}_3}{2m_s+m_Q} ~.
\ea
The reduced masses are given by $m_{\rho}=m_s$ and $m_{\lambda}=3m_sm_Q/(2m_s+m_Q)$, and  
the frequencies of the oscillators in the $\rho$ and the $\lambda$ coordinate by 
$\omega_{\rho}=\sqrt{3C/m_{\rho}}$ and $\omega_{\lambda}=\sqrt{3C/m_{\lambda}}$. 
For equal masses the two frequencies become the same, $\omega_{\rho} = \omega_{\lambda}$.  
For the case of interest in this contribution, $qqQ$ baryons with two light and one heavy 
quark, the freqencies satisfy $\omega_{\lambda} < \omega_{\rho}$. For $QQq$ baryons the situation 
is reversed, $\omega_{\rho} < \omega_{\lambda}$. 

\begin{figure}
\begin{center}
\includegraphics[width=0.6\linewidth]{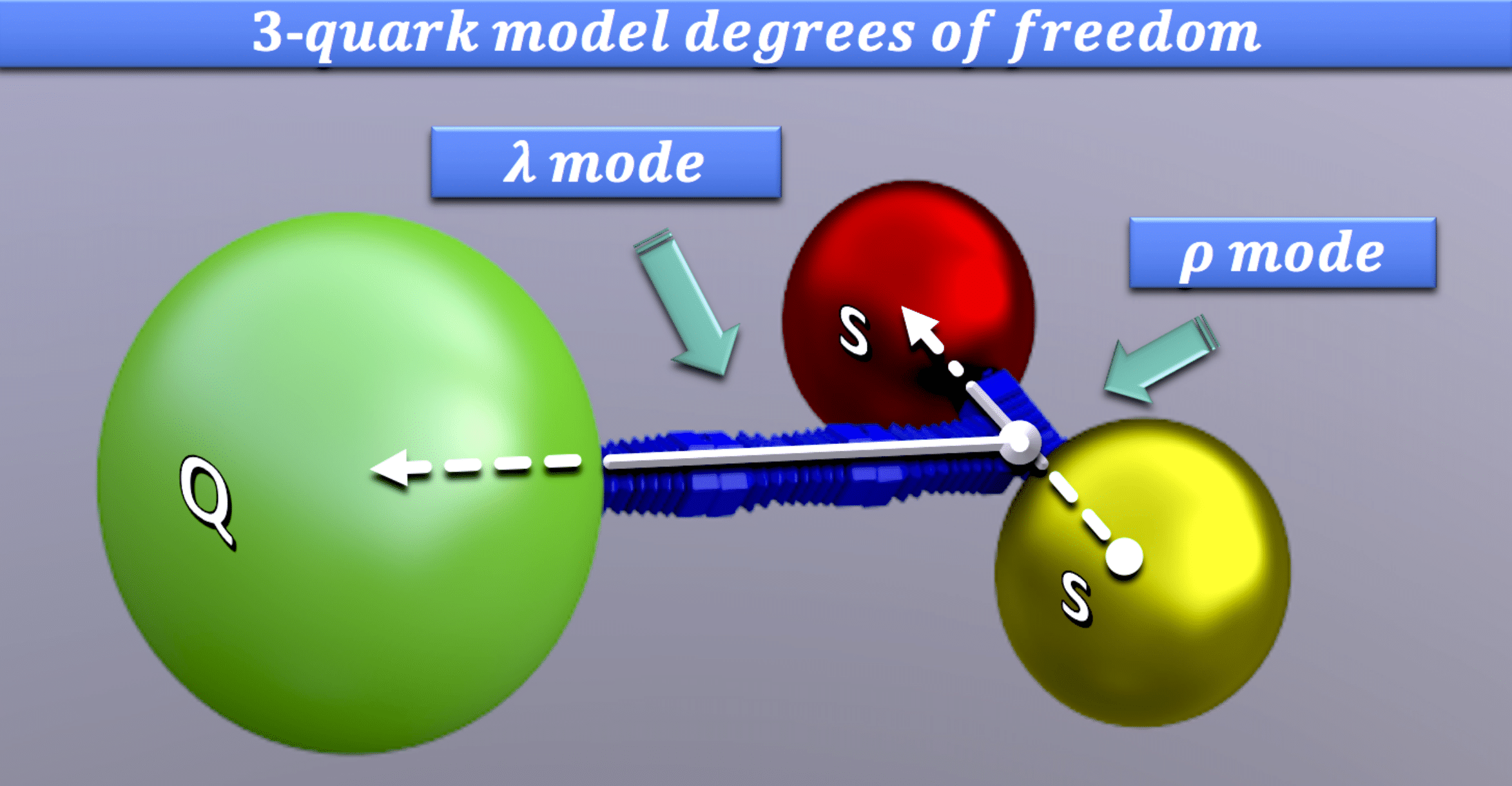}
\caption{Excitation modes, $\rho$ and $\lambda$.}
\label{mode}
\end{center}
\end{figure}

The parameters in the Hamiltonian of Eq.~(\ref{hosc}) were obtained by studying the 
single-charm baryons, $\Sigma_c$, $\Lambda_c$ and $\Xi_c$,  and the single-bottom baryons, 
$\Sigma_b$ and $\Lambda_b$ \cite{epjc}. The results are given in Table~\ref{parameters}. 
We note, that the $ssQ$ configurations in Table~\ref{ssQ} all have the same isospin $I=0$ 
and all belong to the flavor sextet. Moreover, the sum of the quark masses, $2m_s+m_Q$, 
is the same for all $\Omega_Q$ configurations. As a result, the mass differences only depend 
on the harmonic oscillator frequencies, $\hbar \omega_{\rho}$ and $\hbar \omega_{\lambda}$, 
the spin term $A$ and the spin-orbit term $B$
\ba
M &=& M( ^{2}\Omega_Q) + \hbar \omega_{\rho} n_{\rho} + \hbar \omega_{\lambda} n_{\lambda} 
+ A \left[S(S+1)-\frac{3}{4}\right] 
\nonumber\\
&& + B \frac{1}{2} \left[ J(J+1) - L(L+1) - S(S+1) \right] ~.
\ea 
Here $L$, $S$ and $J$ represent the orbital angular momentum, the spin and the total 
angular momentum, respectively. 

\begin{table}
\centering
\caption[]{Parameter values for $\Omega_c$ and $\Omega_b$ baryons 
($ssQ$ baryons with $Q=c$ and $Q=b$, respectively) \cite{epjc}.}
\label{parameters}
\vspace{10pt}
\begin{tabular}{cccl}
\hline
\noalign{\smallskip}
& $Q=c$ & $Q=b$ & \\
\noalign{\smallskip}
\hline
\noalign{\smallskip}
$m_s$ & $450$  & $450$  & MeV \\
$m_Q$ & $1605$ & $4920$ & MeV \\
$C$ & $0.0328$ & $0.0235$ & GeV$^3$ \\
$A$ & $21.54 \pm 0.37$ & $ 6.73 \pm 1.63$ & MeV \\
$B$ & $23.91 \pm 0.31$ & $ 5.15 \pm 0.33$ & MeV \\
$E$ & $30.34 \pm 0.23$ & $26.00 \pm 1.80$ & MeV \\
$G$ & $54.37 \pm 0.58$ & $70.91 \pm 0.49$ & MeV \\
\noalign{\smallskip}
\hline
\end{tabular}
\end{table}

In this contribution we only consider $\Omega_c$ and $\Omega_b$ baryons associated with the 
ground-state configuration $\psi_0$ with $(n_{\rho},n_{\lambda})=(0,0)$, and with one quantum 
of excitation, either in the $\lambda$ mode, $\psi_{\lambda}$ with $(n_{\rho},n_{\lambda})=(0,1)$, 
or in the $\rho$ mode, $\psi_{\rho}$ with $(n_{\rho},n_{\lambda})=(1,0)$ (see Fig.~\ref{mode}). 
The allowed $ssQ$ configurations are given in Table~\ref{ssQ}. 
The ground state $\psi_0$ can be combined with spin $S=1/2$ or $3/2$ to give two positive parity 
baryons, $^{2}\Omega_Q$ and $^{4}\Omega_Q$, with $J^P=1/2^+$ and $3/2^+$, respectively. 
The $\lambda$ excitation gives rise to a total of five negative parity baryon resonances, 
whereas the $\rho$ excitation leads to two negative parity states.

\begin{table}[h]
\centering
\caption[]{Classification of $ssQ$ baryons.}
\label{ssQ}
\vspace{10pt}
\begin{tabular}{clccc}
\hline
\noalign{\smallskip}
State & Wave function & $(n_{\rho},n_{\lambda})$ & $L$ & $J^P$ \\
\noalign{\smallskip}
\hline
\noalign{\smallskip}
$^{2}\Omega_Q$ & $ssQ \left[ \psi_{0} \times \chi_{\lambda} \right]$ 
& $(0,0)$ & $0$ & $\frac{1}{2}^+$ \\ 
\noalign{\smallskip}
$^{4}\Omega_Q$ & $ssQ \left[ \psi_{0} \times \chi_{S} \right]$ 
& $(0,0)$ & $0$ & $\frac{3}{2}^+$ \\ 
\noalign{\smallskip}
$^{2}\lambda_J(\Omega_Q)$ & $ssQ \left[ \psi_{\lambda} \times \chi_{\lambda} \right]$ 
& $(0,1)$ & $1$ & $\frac{1}{2}^-$, $\frac{3}{2}^-$ \\ 
\noalign{\smallskip}
$^{4}\lambda_J(\Omega_Q)$ & $ssQ \left[ \psi_{\lambda} \times \chi_{S} \right]$  
& $(0,1)$ & $1$ & $\frac{1}{2}^-$, $\frac{3}{2}^-$, $\frac{5}{2}^-$ \\ 
\noalign{\smallskip}
$^{2}\rho_J(\Omega_Q)$ & $ssQ \left[ \psi_{\rho} \times \chi_{\rho} \right]$  
& $(1,0)$ & $1$ & $\frac{1}{2}^-$, $\frac{3}{2}^-$ \\ 
\noalign{\smallskip}
\hline
\end{tabular}
\end{table}

\section{Masses and decay widths}

\begin{figure}
\centering
\includegraphics[width=11cm]{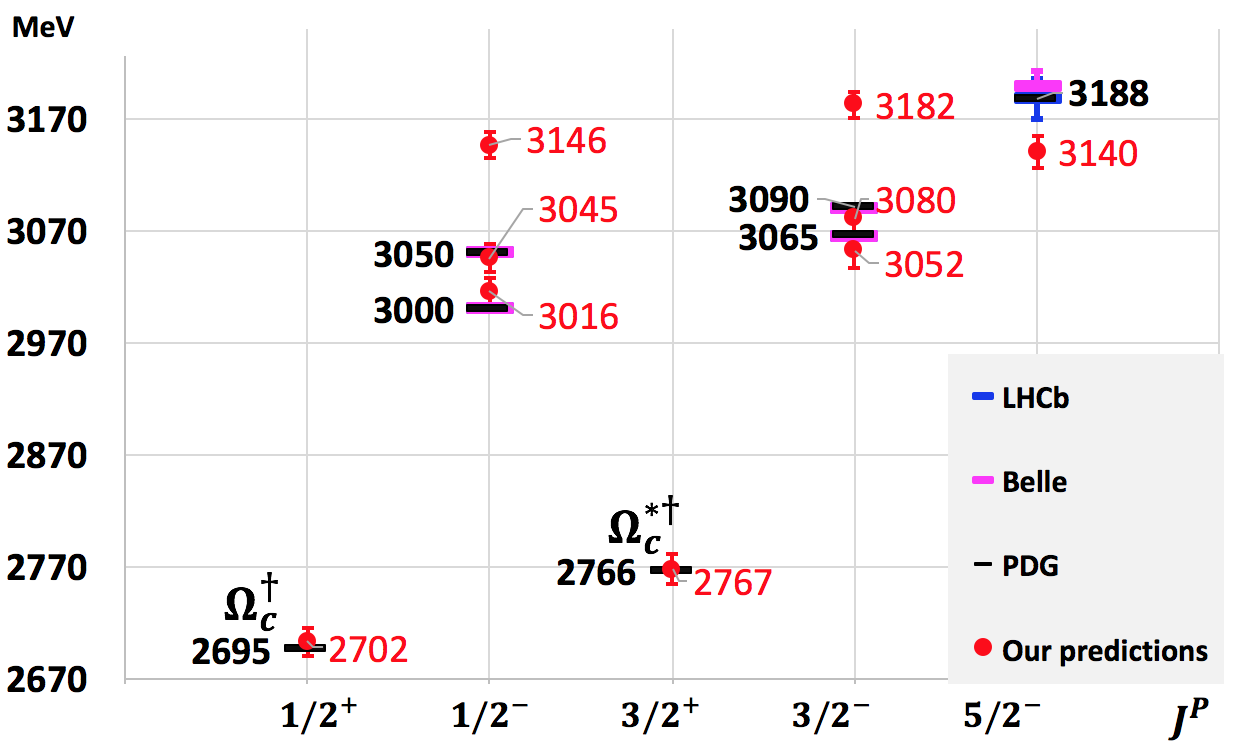}
\caption{$\Omega_c$ mass spectrum with tentative quantum number assignments. Quark model \cite{epjc} 
(red dots), LHCb \cite{Omegac} (blue line), Belle \cite{Belle} (violet line) and PDG \cite{PDG} 
(black line).}
\label{massc}
\end{figure}  

In Fig.~\ref{massc} we show a comparison of the theoretical (red dots) and experimental mass spectrum 
(blue, violet and black lines) of $\Omega_c$ baryons in combination with a tentative assignment of 
quantum numbers. The $\Omega_c(3000)$ and $\Omega_c(3065)$ baryons are assigned to the $\lambda$-mode 
excitation $^{2}\lambda_J(\Omega_Q)$ configuration with $J^P=1/2^-$ and $3/2^-$, respectively, 
and the $\Omega_c(3050)$, $\Omega_c(3090)$ and $\Omega_c(3188)$ baryons to the $\lambda$-mode excitation 
$^{4}\lambda_J(\Omega_Q)$ configuration with $J=1/2^-$, $3/2^-$ and $5/2^-$, respectively.  
We find a good agreement between the theoretical mass spectrum and the experimental data. 
With the exception of the lightest and the heaviest resonant states, $\Omega_c(3000)$ and $\Omega_c(3188)$, 
respectively, the theoretical masses are in agreement with the data within the experimental error, which 
is very small (less than 1 MeV). 

Similarly to Refs.~\cite{Huang:2018wgr,Nieves:2017jjx,Debastiani:2017ewu}, we suggest a molecular 
interpretation of the $\Omega_c(3119)$ state which was observed by the LHCb \cite{Omegac}, but not by 
Belle \cite{Belle}.

\begin{figure}[b]
\centering
\includegraphics[width=11cm]{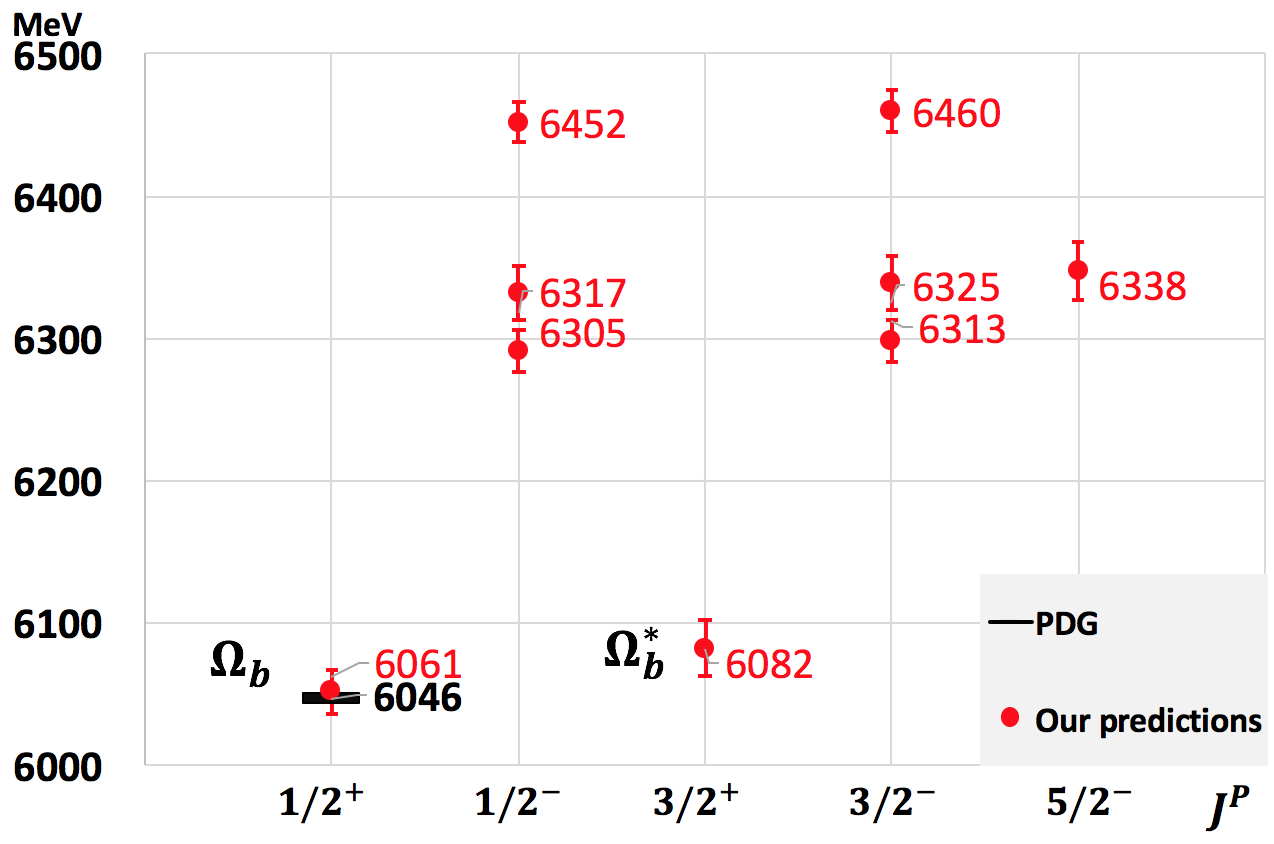}
\caption{Predicted $\Omega_b$ mass spectrum.  
Quark model \cite{epjc} (red dots) and PDG \cite{PDG} (black line).}
\label{massb}
\end{figure}  

Until recently, the experimental knowledge on $\Omega_b$ baryons was limited to $\Omega_{\rm b}^-$ 
with mass $6046$ MeV \cite{PDG} which we assign as the $^{2}\Omega_b$ quark model state.   
In Fig.~\ref{massb}, we show the results of our calculations for the ground- and excited-state 
$\Omega_b$ baryons \cite{epjc}. Shortly thereafter, the LHCb Collaboration announced 
the first observation of excited $\Omega_b^-$ states as four peaks in the $\Xi_b^0 K^-$ 
mass spectrum \cite{Omegab} with decay widths of the order of a few MeV. In Table~\ref{omegacb} we 
make a comparison between our predictions and the new experimental results. The assignment of quantum 
numbers is based on the masses.

In Table~\ref{omegacb} we also show the results for the strong decays  
\ba
\Gamma(ssc \rightarrow qsc + s\overline{q}) &=& \Gamma(\Omega_c \rightarrow \Xi_c + \bar{K}) 
+ \Gamma(\Omega_c \rightarrow \Xi'_c + \bar{K}) ~, 
\nonumber\\
\Gamma(ssb \rightarrow qsb + s\overline{q}) &=& \Gamma(\Omega_b \rightarrow \Xi_b + \bar{K}) 
+ \Gamma(\Omega_b \rightarrow \Xi'_b + \bar{K}) ~.
\ea
The theoretical values were calculated in a $^3P_0$ decay model in which the strength parameter 
was fitted to the decay of the $\Omega_c(3065)$ baryon. The experimental data correspond to the 
total decay width, there is no information on branching ratios. Table~\ref{omegacb} shows 
that the theoretical widths which are based on the mass estimates and the quantum number assignments, 
are compatible with the present experimental data for both the $\Omega_c$ and $\Omega_b$ baryons. 
In particular, the $\lambda$-mode decay widths of the $\Omega_c$ and $\Omega_b$ states are of the 
order of a few MeV, whereas the $\rho$-mode decay widths are forbidden by selection rules. 

The LHCb Collaboration reported the observation of four excited $\Omega_b$ baryons \cite{epjc}. 
In a quark model analysis like the present one, one would expect, just as for the $\Omega_c$ baryons, 
the occurrence of five such states as $\lambda$-mode excitations. The $\rho$-mode excitations arise at 
a somewhat higher mass, but they are decoupled from the $\Xi_b^0 K^-$ decay channel. 

\begin{table}
\centering
\caption{Masses and decay widths of $\Omega_c$ (top) and $\Omega_b$ baryons (bottom). 
Experimental data taken from \cite{PDG,Omegac,Omegab}.}
\vspace{10pt}
\label{omegacb}
\begin{tabular}{cccccc}
\hline
\noalign{\smallskip}
State & $M$ (MeV) & $\Gamma$ (MeV) & Exp & Mass (MeV) & $\Gamma_{\rm tot}$ (MeV) \\
\noalign{\smallskip}
\hline
\noalign{\smallskip}
$^{2}\Omega_c$ & $2702 \pm 12$ & $-$ & $\Omega_c$ & $2695.2 \pm 1.7$ & \\
\noalign{\smallskip}
$^{4}\Omega_c$ & $2767 \pm 13$ & $-$ & $\Omega_c(2770)$ & $2765.9 \pm 2.0$ & \\
\noalign{\smallskip}
$^{2}\lambda(\Omega_c)_{1/2}$ & $3016 \pm  9$ & $0.48$ 
& $\Omega_c(3000)$ & $3000.4 \pm 0.2$ & $4.5 \pm 0.6 \pm 0.3$ \\
\noalign{\smallskip}
$^{4}\lambda(\Omega_c)_{1/2}$ & $3045 \pm 13$ & $1.0$  
& $\Omega_c(3050)$ & $3050.2 \pm 0.1$ & $< 1.2$ \\
\noalign{\smallskip}
$^{2}\lambda(\Omega_c)_{3/2}$ & $3052 \pm 15$ & $3.5^{\ast}$   
& $\Omega_c(3065)$ & $3065.5 \pm 0.3$ & $3.5 \pm 0.4 \pm 0.2$ \\
\noalign{\smallskip}
$^{4}\lambda(\Omega_c)_{3/2}$ & $3080 \pm 13$ & $1.09$ 
& $\Omega_c(3090)$ & $3090.0 \pm 0.5$ & $8.7 \pm 1.0 \pm 0.8$ \\
\noalign{\smallskip}
Molecule &&& $\Omega_c(3120)$ & $3119.1 \pm 1.0$ & $< 2.6$ \\
\noalign{\smallskip}
$^{4}\lambda(\Omega_c)_{5/2}$ & $3140 \pm 14$ & $9.87$  
& $\Omega_c(3188)$ & $3188 \pm 14$ & $60 \pm 26$ \\
\noalign{\smallskip}
$^{2}\rho(\Omega_c)_{1/2}$ & $3146 \pm 12$ & $0$ & n.o. & & \\
\noalign{\smallskip}
$^{2}\rho(\Omega_c)_{3/2}$ & $3182 \pm 12$ & $0$ & n.o. & & \\
\noalign{\smallskip}
\hline
\noalign{\smallskip}
$^{2}\Omega_b$ & $6061 \pm 15$ & $-$ & $\Omega_{\rm b}$ & $6046.1 \pm 0.7$ & \\
\noalign{\smallskip}
$^{4}\Omega_b$ & $6082 \pm 20$ & $-$ & n.o. & & \\
\noalign{\smallskip}
$^{2}\lambda(\Omega_b)_{1/2}$ & $6305 \pm 15$ & $0.50$ 
& $\Omega_b(6316)$ & $6315.6 \pm 0.6$ & $< 2.8 \pm 4.2$ \\
\noalign{\smallskip}
$^{2}\lambda(\Omega_b)_{3/2}$ & $6313 \pm 15$ & $1.14$   
& $\Omega_b(6330)$ & $6330.3 \pm 0.6$ & $< 3.1 \pm 4.7$ \\
\noalign{\smallskip}
$^{4}\lambda(\Omega_b)_{1/2}$ & $6317 \pm 19$ & $2.79$  
& $\Omega_b(6340)$ & $6339.7 \pm 0.6$ & $< 1.5 \pm 1.8$ \\
\noalign{\smallskip}
$^{4}\lambda(\Omega_b)_{3/2}$ & $6325 \pm 19$ & $0.62$ 
& $\Omega_b(6350)$ & $6349.9 \pm 0.6$ & $< 2.8 \pm 3.2$ \\
\noalign{\smallskip}
$^{4}\lambda(\Omega_b)_{5/2}$ & $6338 \pm 20$ & $4.28$  
& n.o. & & \\
\noalign{\smallskip}
$^{2}\rho(\Omega_b)_{1/2}$ & $6452 \pm 15$ & $0$ & n.o. & & \\
\noalign{\smallskip}
$^{2}\rho(\Omega_b)_{3/2}$ & $6460 \pm 15$ & $0$ & n.o. & & \\
\noalign{\smallskip}
\hline
\end{tabular}
\end{table}

\section{Summary and conclusions}

In this contribution we showed that the five excited $\Omega_c^0$ states observed by the LHCb 
Collaboration \cite{Omegac} and later confirmed by the Belle Collaboration \cite{Belle} can 
be understood in a very simple manner in the quark model as $\lambda$-mode excited states. 
In addition, one expects two more $\Omega_c^0$ states as $\rho$-mode excited states, but these 
have a bit higher mass and, most importantly, are decoupled from the $\Xi_c^+ K^-$ 
decay channel by selection rules. 

The same observations hold for the beauty (or bottom) sector. In the quark model, one predicts the 
occurrence of five $\Omega_b^-$ states as $\lambda$-mode excitations, and two more as $\rho$-mode 
excitations. Meanwhile, four excited $\Omega_b^-$ states have been observed by the LHCb Collaboration 
\cite{Omegab} with masses very close to the predicted values. 

Our conclusions are not only based on mass systematics, but also on decay widths. The strong decay 
widths of the channels $\Omega_c^0 \rightarrow \Xi_c^+ K^-$ and $\Omega_b^- \rightarrow \Xi_b^0 K^-$ 
were calculated in a $^3P_0$ model to be of the order of a few MeV, in agreement with the experimental 
data for the total widths. Since there is no experimental information available on branching ratios, 
the only constraint we have is that the calculated widths have to be smaller than the experimental values  
for the total widhs. 

A final remark concerns the quantum numbers of the $\Omega_c^0$ and $\Omega_b^-$ states. The spin and 
parity assignments in Table~\ref{omegacb} are based on mass systematics and decay widths. Their values 
have not (yet) been determined experimentally.

\ack
This work was supported in part by grant IN101320 from DGAPA-UNAM and by grant 251718 from CONACyT, Mexico.

\end{document}